\documentclass{aa}
\usepackage{graphicx}
\usepackage{txfonts}
\usepackage{natbib}
\bibpunct{(}{)}{;}{a}{}{,} % to follow the A&A style

\newcommand{\D}{{\mathrm d}}

\newcommand{\SBrak}[1]{{\left[{#1}\right]}}

\newcommand{\DxDyInd}[3]{{\Brak{\frac{\D{#1}}{\D{#2}}}_{{#3}}}}
\newcommand{\dxdycz}[3]{{\Brak{\frac{\partial{#1}}{\partial{#2}}}_{{#3}}}}
\newcommand{\Brak}[1]{{\left({#1}\right)}}

 \def\mso{\,\mathrm{M}_\odot}

 \def\kms{\, {\rm km}\, {\rm s}^{-1}}

 \def\simle{\mathrel{\hbox{\rlap{\hbox{\lower4pt\hbox{$\sim$}}}\hbox{$<$}}}}
 \def\simgr{\mathrel{\hbox{\rlap{\hbox{\lower4pt\hbox{$\sim$}}}\hbox{$>$}}}}

 \def\dmag{D_{\mathrm{mag}}}
 \def\grad{\nabla}
 \def\adgrad{\nabla_{\mathrm{\!ad}}}
 \def\mugrad{\nabla_{\!\mu}}
 \def\ath{\alpha_{\mathrm{th}}}

 \def\c2{^{12}{\mathrm C}}
 \def\c3{^{13}{\mathrm C}}
 \def\n14{^{14}{\mathrm N}}
 \def\c1213{^{12}{\mathrm C}/^{13}{\mathrm C}}
 \def\he3he4{^3\mathrm{He}/^4\mathrm{He}}
 \def\he3{^3\mathrm{He}}
 \def\vsini{\varv\sin i}
\begin{document}
   \title{Thermohaline mixing in evolved low-mass stars}

   \author{M. Cantiello
           \and
          N. Langer}
   \offprints{Matteo Cantiello \email{cantiello@astro.uni-bonn.de}}

   \institute{   
             Argelander-Institut f\"ur Astronomie der Universit\"at Bonn, Auf dem H\"ugel 71, 53121 Bonn, Germany
             }

   \date{Received 2010 / Accepted 2010}

  \abstract
  {Thermohaline mixing has recently been proposed to occur in
low-mass red giants, with large consequence for the chemical yields
of low-mass stars.  }
  {We investigate the role of thermohaline mixing during the evolution
of stars between 1$\mso$ and 3$\mso$, in comparison with other mixing processes
acting in these stars.  }
  {We use a stellar evolution code which includes rotational mixing,
internal magnetic fields and thermohaline mixing.  }
  {  We confirm that during the red giant stage, thermohaline mixing has the potential to decrease the abundance of $^3$He, which is produced earlier on the main sequence.
In our models we find that this process is working on the RGB only in stars with initial mass  $\textrm{M} \simle1.5\mso$. 
Moreover we report that thermohaline mixing is also present during core He-burning and beyond, and has the potential to change the surface abundances of AGB stars. 
While we find rotational and magnetic mixing to be negligible compared to
the thermohaline mixing in the relevant layers, the interaction of thermohaline 
motions with the differential rotation may be essential to establish the timescale
of thermohaline mixing in red giants.}
  {  To explain the surface abundances observed at the bump in the luminosity function, the speed of the mixing process needs to be more than two orders of magnitude higher than in our models.
 However it is not clear if thermohaline mixing is the only physical process responsible for these surface-abundance anomalies. Therefore it is not possible at this stage to calibrate the efficiency of thermohaline mixing against the observations.}


   \maketitle

%________________________________________________________________

\section{Introduction}

Stars are rotating, self-gravitating balls of hot plasma. 
Due to thermonuclear reactions in the deep stellar interior,
stars, which presumably start out chemically homogeneous,
develop chemical inhomogeneities.  At the densities
typically achieved in stars, thermal diffusion, or Brownian
motion, is not able to lead to chemical mixing.
However, various turbulent mixing processes are thought to act
inside stars, leading to transport of chemical species, heat, angular
momentum, and magnetic fields \citep{pin97,hlw00,hws05}.

Thermohaline mixing is usually not considered to be an important mixing process
in single stars, because the ashes of thermonuclear fusion consist of heavier nuclei
than its fuel, and stars usually burn from the inside out. The condition for
thermohaline mixing, however, is that the mean molecular weight ($\mu$)
decreases inward. 
This can occur in accreting binaries, and the importance of thermohaline mixing 
has long been recognized by the binary community \citep[e.g.,][]{1989A&A...211..356D,1992MNRAS.259...17S,Wellstein:2001p121}.
Recently \citet[CZ07]{cz07} identified thermohaline mixing 
as an important mixing process, which significantly modifies the surface composition
of red giants after the first dredge-up. The work by CZ07 was initiated by the paper
of \citet[EDL06]{Edl06}, who found an mean molecular weight ($\mu$) 
inversion --- i.e., $\left( d\log\mu \over d\log P \right) < 0$ --- 
below the red giant convective envelope in a 1D-stellar evolution
calculation. While EDL06 then investigated the stability of the zone containing the
$\mu$-inversion with a 3D hydro-code and found these layers to be Rayleigh-Taylor-unstable,
CZ07 could not confirm this, but found the layers to be unstable due to thermohaline
mixing. 

\citet{Edl06} found a $\mu$-inversion in their $1\mso$ stellar evolution model, occurring
after the so-called luminosity bump on the red giant branch, which is produced
after the first dredge-up, when the H-burning shell source enters the
chemically homogeneous part of the envelope. The $\mu$-inversion is produced by the
reaction $^3$He($^3$He,2p)$^4$He, as predicted by \citet{Ulr72}. It does not occur earlier,
because the magnitude of the $\mu$-inversion is small and negligible compared to a stabilizing
$\mu$-stratification.

Mixing processes below the convective envelope in models of low-mass stars turn out
to be essential for the prediction of their chemical yield of $^3$He (EDL06), and
are essential to understand the surface abundances of red giants
--- in particular the $^{12}$C/$^{13}$C ratio, $^7$Li and the carbon and nitrogen
abundances (CZ07). 
This may also be important for other occurrences of thermohaline mixing in stars,
i.e., in single stars when a $\mu$-inversion is produced by off-center ignition
in semi-degenerate cores \citep{2009A&A...497..463S} or in stars which accrete chemically enriched matter 
from a companion in a close binary \citep[e.g.,][]{sgi+07}. Accreted  metal-rich matter
during the phases of planetary formation also leads to thermohaline mixing, which can reconcile the
observed metallicity distribution of the central stars of planetary systems \citep{Vau04}. 

In the present paper we investigate the evolution
of solar metallicity stars between $1\mso$ and $3\mso$ from the ZAMS up to the
thermally-pulsing AGB stage, based on models computed 
during the last years. We show for which initial mass range and during which evolutionary
phase thermohaline mixing occurs and what consequences it has. Besides thermohaline mixing,
our models include
convection, rotation-induced mixing, and internal magnetic fields, and we compare 
the significance of these processes in relation to the thermohaline mixing. 

\section{The speed of thermohaline mixing }
\subsection{Thermohaline mixing VS Rayleigh-Taylor}
As pointed out by CZ07, a $\mu$-inversion inside a star should give rise to thermohaline mixing,
which is a slow mixing process acting on the local thermal timescale. Could a
Rayleigh-Taylor instability be present in these layers? Indeed, EDL06 interprete the
origin of the instability which they find in their 3D models as due to the buoyancy
$ g \left(\Delta\mu \over \mu\right)$ produced by the $\mu$-inversion, i.e. a dynamical effect.
But a dynamical instability should only occur if the $\mu$-inversion were to lead to a
density inversion. However, this would only be possible if the considered layers
were convectively unstable in the hydrostatic 1D stellar evolution models.
As pointed out by CZ07 and as confirmed by our models,
the $\mu$-inversion produced by the $^3$He($^3$He,2p)$^4$He reaction does not induce
convection. We conclude that the Rayleigh-Taylor instability may not be a likely explanation
of the hydrodynamic motions found by EDL06.
A similar conclusion was also reached by \citet{Denissenkov:2008p5212}, who studied in detail
the instability driven by the $\mu$-inversion in both the adiabatic and the radiative limit. 

Because smaller blobs have a smaller thermal timescale \citep{Kippenhahn:1980p5400},
could EDL06 have found the high wavenumber tale of the thermohaline instability?
They found the instability to occur within 2000\,s. The size of a blob with such a
short thermal timescale above the H-burning shell of a red giant is on the
order of 50\,km. This is too small to be resolved in the 3D model shown by ELD06.
Furthermore, an inspection of their Fig.~5 reveals that the length scale of the
instability they found is on the order of $10^3...10^4\,$km, which corresponds to
thermal timescales of about 1~yr. Therefore, it seems unlikely that EDL06
actually picked up the thermohaline instability in their 3D hydrodynamic model,
unless its non-linear manifestation involves a timescale much shorter than the thermal
timescale.

\subsection{The efficiency of thermohaline mixing}\label{efficiency}
In Sect.~\ref{method} we explain the details of our implementation of thermohaline mixing in 1D stellar evolution calculations.
The diffusion coefficient for the mixing process contains a parameter which depends on the geometrical configuration of the fluid elements. 
This parameter ($\ath$) is very important to understand the role of thermohaline mixing. It determines the timescale 
of the mixing (the velocity of the fingers/blobs) that we show in Sect.~\ref{rgb} and \ref{beyond} plays a role not only in determining how fast the surface abundances of
redgiants can change, but also if thermohaline mixing is present in stars of different mass and at different evolutionary phases.

\citet{cz07} showed that a high value of $\ath$ is needed to match the surface abundances of field stars after the luminosity bump \citep{Gratton:2000p5442}.
Similar to the value adopted by \citet{Ulr72} they use an efficiency factor  corresponding to $\ath=667$ in our diffusion coefficient.
This value corresponds to the diffusion process involving fingers with an aspect ratio (length/width) of 5. 
\citet{Denissenkov:2008p5212} claim that to explain the observed mixing pattern in low-mass RGB stars, fluid elements have to travel over length scales exceeding their diameters by a factor of 10 or more.

\begin{figure}
\resizebox{\hsize}{!}{\includegraphics{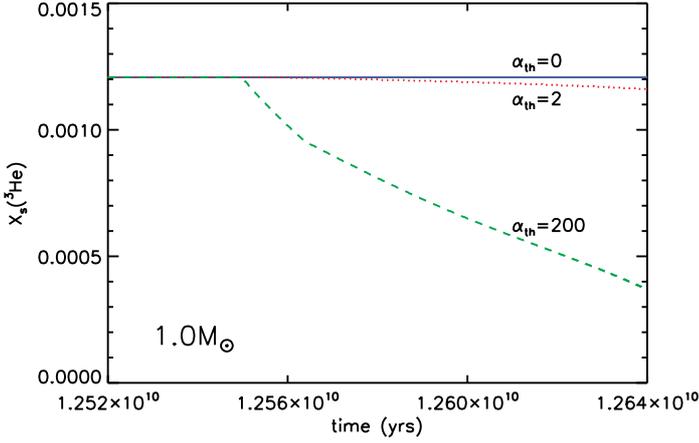}}
 \caption{Evolution of the surface abundance of $^{3}$He for ${\alpha}_{th}=0$
 (blue solid line), ${\alpha}_{th}=2$ (red dotted line) and ${\alpha}_{th}=200$ 
(green dashed line) from before the onset of thermohaline mixing to the core He-flash
 in a $1.0\mso$ star.}
\label{he3surf} 
\end{figure}

On the other hand, the order of magnitude of our efficiency parameter $\ath=2.0$ corresponds to the prescription of \citet{Kippenhahn:1980p5400}, where the 
diffusion process involves  blobs of size L traveling a mean free path L before dissolving. 
The same prescription has been used by \citet{sgi+07}, who dealt with the problem of thermohaline mixing in accreting binaries.
The sensitivity of thermohaline mixing to a change of the efficiency parameter  is shown in Fig.~\ref{he3surf}, where the change of  the surface abundance of $\he3$ 
at the luminosity bump is shown for different values of $\ath$.\\

That a prescription can reproduce the observed surface abundances may not  be sufficient to prefer it over others. 
It is possible that other mixing processes are at work. The resulting observed abundances could still be mainly due to thermohaline mixing, as proposed by CZ07,  
but at this stage it is not possible to exclude that  magnetic buoyancy \citep{Busso:2007p5186,Nordhaus:2008p5185}, or the interaction of different mixing processes 
\citep[e.g., thermohaline mixing and magnetic buoyancy,][]{Denissenkov:2009p5218}, could play an even more important role. 

To clarify the picture here we discuss the main differences between the two physical prescriptions for thermohaline mixing.
Experiments of thermohaline mixing show slender fingers in the linear regime \citep[e.g.,][]{kri03}, supporting the picture of \citet{Ulr72}. On the other
hand the physical conditions inside a star are quite different from those in the laboratory. In particular the Prandtl number $\sigma$,
defined as the ratio of the kinematic viscosity $\nu$ to the thermal diffusivity $\kappa_T$, is very small in stars ($\sigma \sim 10^{-6}$). This number
is about 7 in  water, where most of thermohaline mixing experiments have been performed. The question arises if for such small values of  $\sigma$
a finger-like structure can be stable, especially in layers where shear and horizontal turbulence is present. 2D hydrodynamic simulations of double-diffusive phenomena,  without external perturbations have been performed in the past \citep[e.g.,][]{Merryfield:1995p5448,BascoulPhd,Bascoul:2007p5449}. While the resolution required by the physical conditions in stellar interiors is computationally not accessible, lowering the Prandtl number to values of about $10^{-2}$ always results in increasingly unstable structures \citep{Merryfield:1995p5448,BascoulPhd}.
Therefore it might be dangerous to assume that the same configuration of thermohaline mixing, as observed in water, is occurring in stars.

As we show in Sect.~\ref{rotation}, the radiative buffer between the H-burning shell and the convective envelope is
a region of the star where the angular velocity is rapidly changing. Even if the order of magnitude of the rotationally-induced
instabilities is much lower than the one from thermohaline mixing (cf. Fig \ref{diffusion}), it is possible that the interaction of the shear motions 
with the thermohaline diffusion prevents a relatively ordered flow to be stable, contrary to Ulrich's assumption. 
That shear can decrease the efficiency of thermohaline mixing was already pointed out by \citet{Canuto:1999p5353}. 
The effect of strong horizontal turbulence in stellar layers has also been discussed by \citet{Denissenkov:2008p5212}, and we conclude in Sect.~\ref{interaction} that 
this effect could work against the fingers and in favor of blobs. 

The thermohaline mixing prescription proposed by \citet{Kippenhahn:1980p5400}
 has been used for the calculations presented here.
It is clear from the discussion above that a better calibration of the mixing speed requires realistic hydrodynamic calculations of the instability.
 
\section{Method}\label{method}
We use a 1-D hydrodynamic stellar evolution code \citep[][and references therein]{Yln06}. 
Mixing is treated as a diffusive process and is implemented by solving the diffusion equation
\begin{equation}
  \dxdycz{X_n}{t}{m}=\dxdycz{}{m}{t}\, \SBrak{(4\pi r^2 \rho)^2 \, D \,
     \dxdycz{X_n}{m}{t}}+\DxDyInd{X_n}{t}{\mathrm{nuc}} \label{diff},
\end{equation}
where $D$ is the diffusion coefficient constructed from the sum of
individual mixing processes and $X_n$ the mass fraction of species
$n$.  The second term on the right hand side accounts for nuclear
reactions. The contributions to the diffusion coefficient are convection, semiconvection, 
thermohaline mixing, rotationally induced mixing, and magnetic diffusion.
The code includes the effect of centrifugal force on the stellar structure, and
the transport of angular momentum is also treated as a diffusive process \citep{es78,Pks+89}.

The condition for the occurrance of thermohaline mixing is
\begin{equation}
\frac{\varphi}{\delta} \,\mugrad \le \grad - \adgrad \le 0 \label{condition},
\end{equation}
i.e. the instability operates in regions that are stable against convection (according to the 
Ledoux criterion) and where an inversion in the mean molecular weight is present. Here 
$\varphi =(\partial \ln \rho / \partial \ln
\mu )_{P,T}$, $\delta =-(\partial \ln \rho / \partial \ln T )_{P,\mu
  }$, $\nabla_\mu = d\ln\mu / d\ln P$, $\nabla_{ad}= (\partial \ln T/
\partial \ln P )_{ad}$, and $\nabla = d\ln T / d\ln P$.
Numerically, we treat thermohaline mixing through a diffusion scheme
\citep{bra97,Wellstein:2001p121}. The corresponding diffusion coefficient is based on the
work of \citet{Ste60}, \citet{Ulr72}, and \citet{Kippenhahn:1980p5400}; it
reads
\begin{equation}
  D_{th} = -\ath\; \frac{3K}{2\,\rho\, c_P}\,
  \frac{\frac{\varphi}{\delta}\nabla_\mu}{(\adgrad - \grad)} \label{coefficient}, 
\end{equation}
where $\rho $ is the density, $K=4acT^{3}/(3\kappa\rho )$ the thermal conductivity, and 
$ c_P=(dq/dT)_P$ the specific heat capacity. The quantity
$\ath$ is a efficiency parameter for the thermohaline mixing.
The value of this parameter 
depends on the geometry of the fingers arising from the instability and is still a matter of debate 
\citep{Ulr72,Kippenhahn:1980p5400,cz07,Denissenkov:2008p5212}. 
As explained in Sect.~\ref{efficiency} unless otherwise specified we assume a value $\ath=2.0$   for the efficiency of thermohaline mixing. 
This value roughly  corresponds to the prescription of \citet{Kippenhahn:1980p5400}, in which fluid elements (blobs) travel over length scales 
comparable to their diameter.\\
For rotational mixing, four different diffusion coefficients are calculated for dynamical shear, 
secular shear, Eddington-Sweet circulation and Goldreich-Schubert-Fricke instability. 
Details on the physics of these instabilities and their implementation in the code can be found in \citet{hlw00}.
 
Chemical mixing and transport of angular momentum due to magnetic fields \citep{Spruit:2002p36} is included as in \citet{hws05}. 
The contribution of magnetic fields to the mixing is also calculated as a diffusion coefficient ($\dmag$), which is added to the total diffusion
coefficient $D$ that enters Eq.~\ref{diff}.

We compute evolutionary models of $1.0\mso$, $1.5\mso$, $2.0\mso$ and $3.0\mso$ at solar metallicity (Z=0.02).
The initial equatorial velocities of these models were chosen to be 10, 45, 140 and 250 $\kms$ \citep{Tas00}; we assume
the stars to be rigidly rotating at the zero-age main sequence. Throughout the evolution of all models, the mass-loss rate of 
\citet{rei75} was used.  

 \begin{figure}
\resizebox{\hsize}{!}{\includegraphics{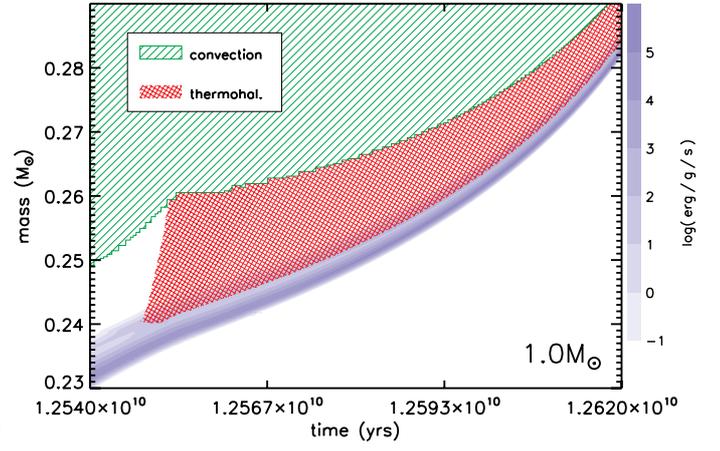}}
\caption{Evolution of the region between the H-burning shell source and the 
convective envelope in the RGB phase after the onset of thermohaline mixing for a $1.0\mso$ star. 
Green hatched regions indicate convection and red cross hatched regions indicate thermohaline mixing, 
as displayed in the legend. Blue shading shows regions of nuclear energy generation, tracing the H-burning shell.} 
\label{zoomhaline} 
\end{figure}

\begin{figure}
\resizebox{\hsize}{!}{\includegraphics{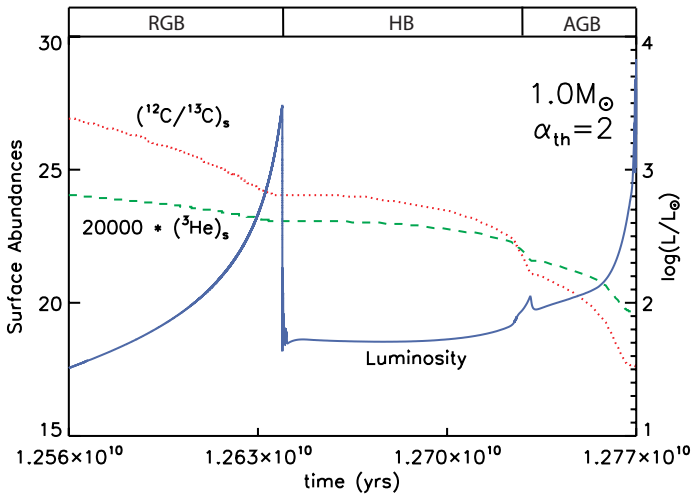}}
 \caption{Evolution of the surface abundance profiles of the $^{12}$C/$^{13}$C ratio 
(dotted red line) and $^3$He (dashed green line), and of the luminosity (solid blue line)
 from the onset of thermohaline mixing up to the AGB for a $1.0\mso$ star.}
\label{surfalles} 
\end{figure}

\section{Thermohaline mixing on the giant branch}\label{rgb}
We compute the stellar models of 1.0, 1.5, 2.0 and 3.0 $\mso$ with solar metallicity. 
The evolutionary calculations presented are the same as in  \citet{2008A&A...481L..87S}, to which we refer for the details of their 
main sequence evolution.

The surface composition of low-mass stars is substantially changed during the first dredge-up: lithium and 
carbon abundances as well as the carbon isotopic ratio decline,$^3$He and nitrogen abundances increase.
 After the first dredge-up the H-burning shell is advancing while the convective envelope retreats;
 the shell source then  enters the chemically homogeneous part of the envelope. \citet{Edl06} and CZ07  have shown 
 that in this situation an inversion of the molecular weight is created by
 the  reaction $^3$He($^3$He,2p)$^4$He in the outer wing of the H-burning shell in  models of 1.0 
and 0.9 $\mso$.  This inversion was already predicted by \citet{Ulr72}.

We confirm an inversion in the mean molecular weight in
the outer wing of the H-burning shell. This inversion occurs after the luminosity bump on the red giant branch in the 1.0, 1.5 and 2.0 $\mso$  models.
The size of the $\mu$-inversion  depends on the local amount of $^3$He and in the studied mass range 
decreases with increasing initial mass\footnote{During the main sequence of these stars, the pp chain operates partially burning hydrogen to $^3$He,
but not beyond, into a wide zone outside the main energy-producing region. At the end of the core H-burning the first dredge-up mixes this $^3$He with the stellar envelope.
Because the main sequence lifetime is longer for lower mass stars, these are able to produce bigger amounts of $^3$He.}.
According to Inequality~\ref{condition} this inversion causes thermohaline mixing in the radiative 
buffer layer, the radiative region between the H-burning shell and the convective envelope.
We emphasize that the extension of the region in which the mixing process is active is not chosen arbitrarily, but is calculated self-consistently by the code. This is done at each 
time step of the evolutionary calculation by checking which grid points fulfill  condition~\ref{condition}. 
This is a major difference between models including thermohaline mixing and models where the extra mixing 
is provided by magnetic buoyancy \citep{Busso:2007p5186,Nordhaus:2008p5185,Denissenkov:2009p5218}.
Indeed for the latter a self-consistent implementation is still not available in 1D 
stellar evolution codes, and the extension of the extra mixing has to be set arbitrarily.

In our 1$\mso$ model thermohaline mixing develops at the luminosity bump and transports chemical species between  
the H-burning shell and the convective envelope (see  Fig.~\ref{zoomhaline}). This results 
in a change of the stellar surface abundances. Fig.~\ref{surfalles} shows the evolution of the $^3$He surface-abundance
and of the ratio $\c1213$, qualitatively confirming the result of EDL06 and CZ07, namely that thermohaline mixing
is depleting $^3$He and lowering the ratio $\c1213$ on the giant branch. 
As already observed by CZ07, the surface abundance of $^{16}$O 
is not affected because thermohaline mixing does not transport chemical species deep enough the H-burning shell.

\begin{figure}
\resizebox{\hsize}{!}{\includegraphics{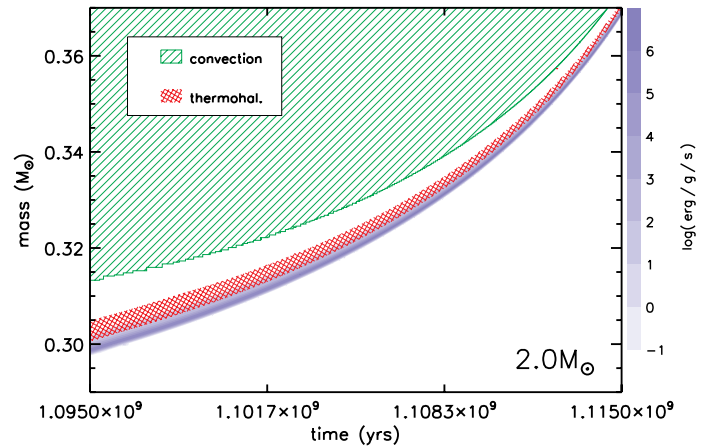}}
\caption{Evolution of the region between the H-burning shell source and the 
convective envelope in the RGB phase after the onset of thermohaline mixing for a $2.0\mso$ star. Green hatched regions indicate convection
 and red cross hatched regions indicate thermohaline mixing, 
as displayed in the legend. Blue shading shows regions of nuclear energy generation. }
 \label{disconnected} 
\end{figure}

Unlike the 1.0$\mso$ and the 1.5$\mso$ model, in the 2.0$\mso$ model thermohaline mixing starts but never connects the H-burning shell
to the convective envelope. This is a direct consequence of the lower $^3$He abundance, which results in smaller $\mu$-inversion and therefore in a slower
thermohaline mixing, according to Eq.~\ref{coefficient}. 
It is surprising that thermohaline mixing, once started in the outer wing of the H-burning shell, does not 
spread through the whole radiative buffer layer.
In fact the H-shell burns in a chemically homogeneous region, meaning that no compositional barrier is expected to stop the instability. 
The reason is that  the region unstable to thermohaline mixing  moves too slowly in the mass coordinates and never catches-up with the quicker receding envelope. 
This situation is shown in  Fig.~\ref{disconnected}. As a result no change in the stellar surface composition due to 
thermohaline mixing is observed during the RGB phase of the 2.0$\mso$ model.

In our 3.0$\mso$ model the H-burning shell never penetrates the homogeneous region left by the 1DUP. 
Accordingly thermohaline mixing does not occur during this phase.

In conclusion our models predict that before the He-core burning thermohaline mixing is able to change surface abundances only in stars with $M \simle 1.5\mso$. 

During the RGB evolution the choice of $\ath=2.0$, roughly corresponding to the prescription of \citet{Kippenhahn:1980p5400} for the thermohaline mixing, allows our stellar models to reach helium ignition without having depleted too much $^3$He in the envelope. The presence of leftover $^3$He allows thermohaline mixing to play a role also during a later evolutionary phase, as we show below.

\begin{figure}
\resizebox{\hsize}{!}{\includegraphics{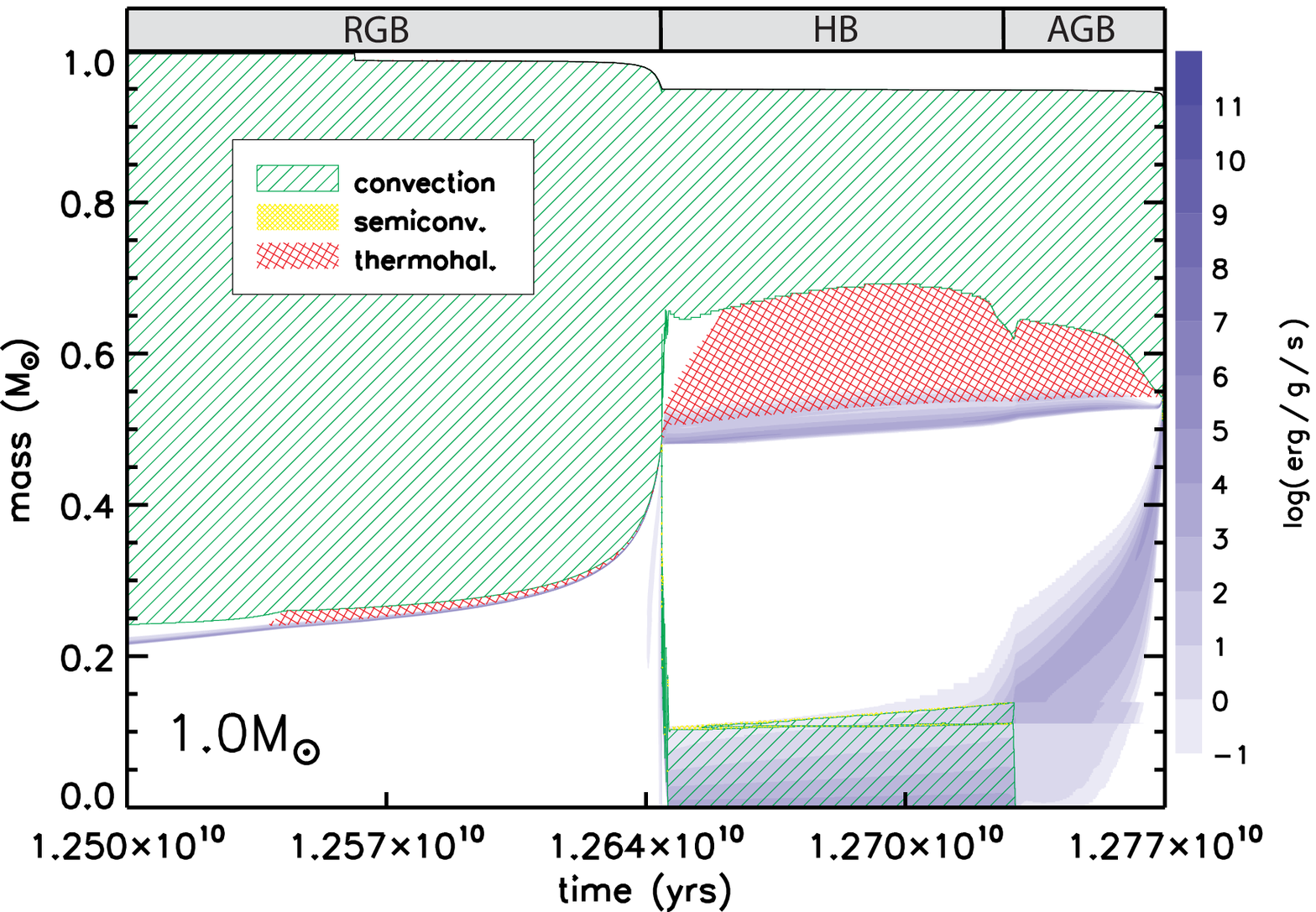}}
 \caption{Evolution of the internal structure of a $1.0\mso$ star from the onset of thermohaline 
mixing to the asymptotic giant branch. Green hatched regions indicate convection, yellow
 filled regions represent semiconvection and red cross hatched regions indicate thermohaline mixing, 
as displayed in the legend. Blue shading shows regions of nuclear energy generation.}
\label{1.0} 
\end{figure}
 
\begin{figure}
\resizebox{\hsize}{!}{\includegraphics{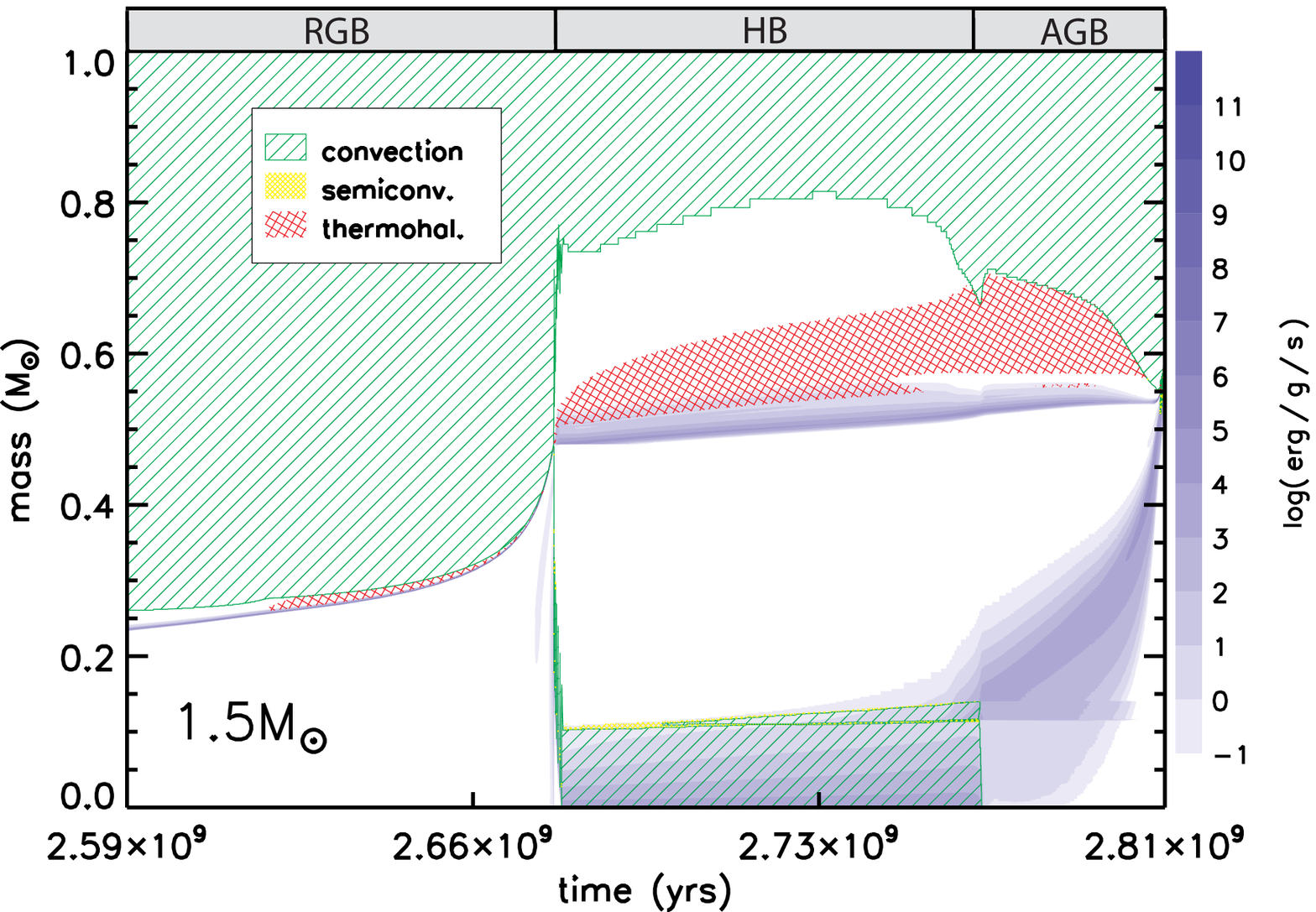}}
 \caption{Evolution of the internal structure of a $1.5\mso$ star from the onset of 
thermohaline mixing to the AGB phase. Green hatched regions 
indicate convection, yellow regions represent semiconvection and regions of thermohaline 
mixing are red cross hatched, as is displayed in the legend. Blue shading shows regions of 
nuclear energy generation.}
\label{1.5} 
\end{figure}

\begin{figure}
\resizebox{\hsize}{!}{\includegraphics{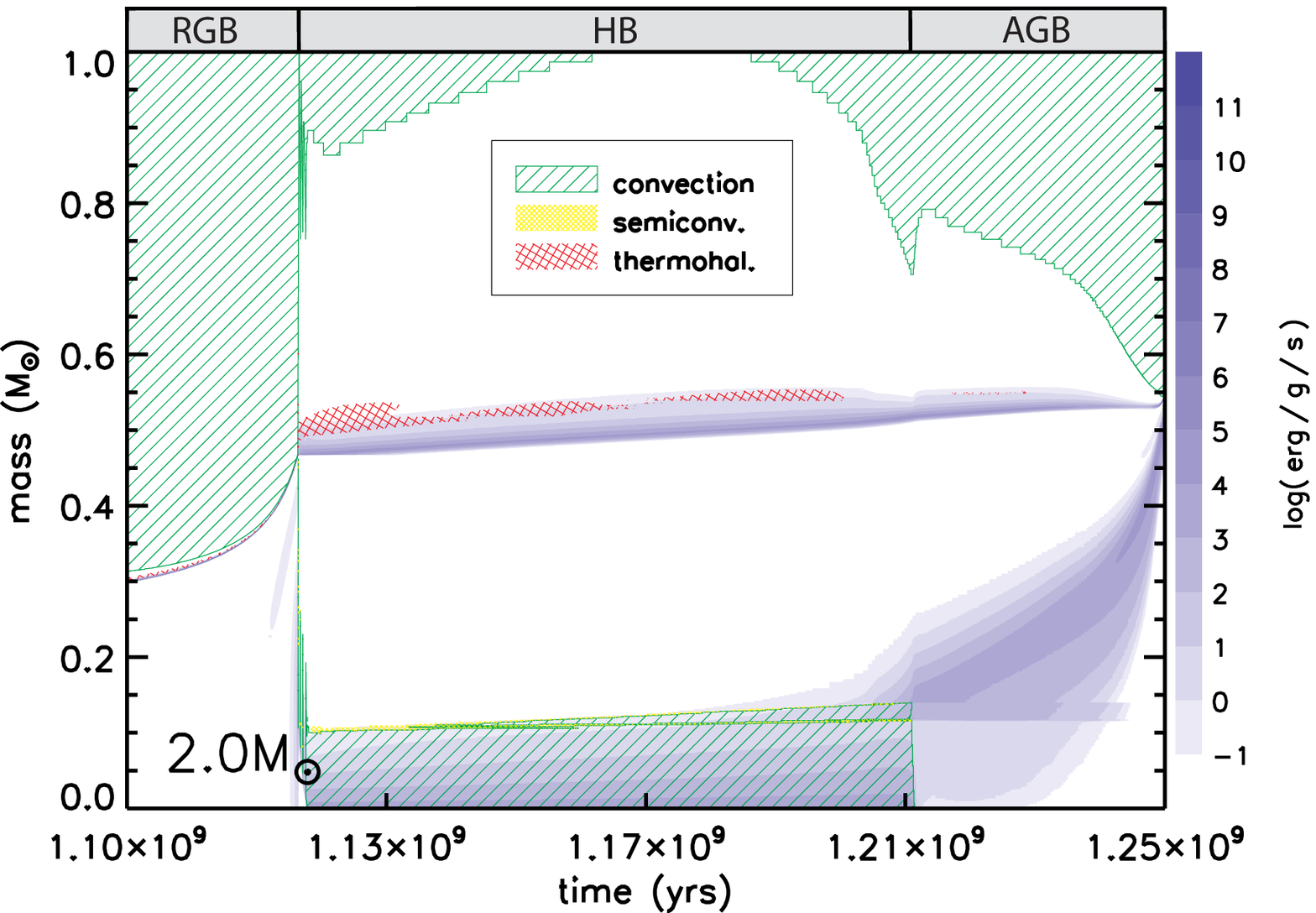}}
 \caption{Evolution of the internal structure of a $2.0\mso$ star from the onset of 
thermohaline mixing to the AGB phase. Green hatched regions 
indicate convection, yellow regions represent semiconvection and regions of thermohaline 
mixing are red cross hatched, as is displayed in the legend. Blue shading shows regions of 
nuclear energy generation.}
\label{2.0} 
\end{figure} 
 
\begin{figure}
\resizebox{\hsize}{!}{\includegraphics{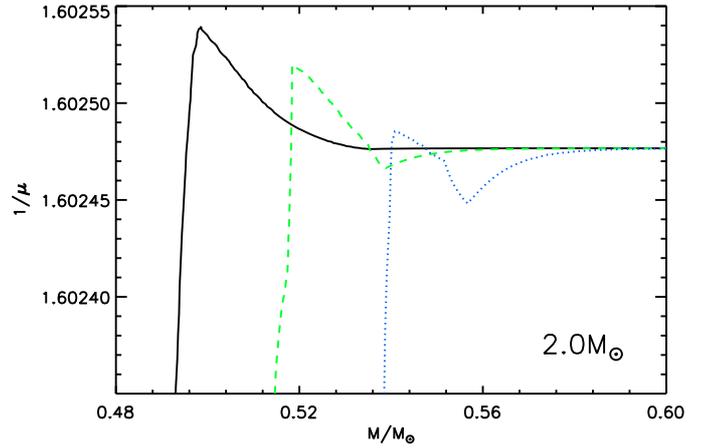}}
 \caption{Profiles of the reciprocal mean molecular weight ($1/\mu$) in the region above the H-burning shell. 
The plot shows three successive times in a 2$\mso$ model during the horizontal branch. The black, continuous line 
represents the model at t $=1.13\times10^{9}$; the green, dashed line shows the same model at t $=1.16\times10^{9}$, while 
the blue, dotted line is the $1/\mu$ profile at t $=1.21\times10^{9}$. }
\label{peaks} 
\end{figure}

\section{Thermohaline mixing on the horizontal branch and during the AGB stage}\label{beyond}  
While CZ07 and EDL07 investigate thermohaline mixing only during the RGB, we followed the evolution of 
our models until the thermally-pulsing AGB stage (TP-AGB).
Indeed a $\mu$-inversion  is always created if a H-burning shell is active in a chemically homogeneous layer, the size of the inversion depending on the local abundance of $^3$He. 
This  happens not only during the RGB, but also during the horizontal branch (HB) and the AGB phase.  As a result  thermohaline mixing can also operate during these evolutionary phases \citep{2008IAUS..252..103C,2010MNRAS.403..505S}.    

Depending on the efficiency of thermohaline mixing during the RGB, the $\he3$ can be exhausted at the end of this phase (e.g in the models of CZ07).
However, stars that avoid extra mixing during the RGB are observed \citep{Cdn98}.  For these stars the $^3$He reservoir is intact at He ignition, and thermohaline mixing has the potential to play an important role during the HB and AGB phases. This is confirmed by the evolutionary calculations presented in Sect.~\ref{hb} and \ref{agb}.

\subsection{Horizontal branch}\label{hb}
After the core He-flash, helium is burned in the core, while a H-burning shell is still active below the convective envelope. 
 In our 1$\mso$ model we found that during this phase thermohaline mixing is present and can spread through the whole radiative buffer layer. This is clear in Fig.~\ref{1.0}
where thermohaline mixing (red, cross hatched region) extends from the H-shell to the convective envelope also after ignition of the core He-burning (HB label in the plot). 
Accordingly surface abundances change during this phase, as shown in Fig.~\ref{surfalles}.  Here a change of surface abundances is 
also visible after the luminosity peak corresponding to the core He-flash.

Contrary to the 1$\mso$ model, in our 1.5 and 2.0$\mso$ models thermohaline mixing does not change the surface abundances during the HB phase. 
In the 1.5$\mso$ model the instability succeeds in connecting the H-shell and the convective envelope only at the end of the core He-burning (Fig.~\ref{1.5}), while in the 2.0$\mso$ model
this is never achieved (Fig.~\ref{2.0}).
In the latter case thermohaline diffusion is confined to a tiny layer 
on top of the H-burning shell, never spreading through the radiative layer (the red cross-hatched region in  Fig.~\ref{2.0}). 
This is due to a $\mu$-barrier, which stops the development of the instability. 
In Fig.~\ref{peaks} we show the profile of $1/\mu$ for the 2.0$\mso$ model at three 
successive times during core He-burning: the initial peak created by the reaction  $^3$He($^3$He,2p)$^4$He  gets smaller, while a dip begins to be visible at slightly 
higher mass coordinate, i.e. at a lower temperature. This $\mu$-barrier is responsible for stopping the instability;  this process is discussed in greater detail in Appendix~A.

In the 3.0$\mso$ model the H-burning shell enters for the first time the chemically homogeneous region after igniting He in the core. However, in this case also
thermohaline mixing does not change the surface abundances because is not able to connect the H-burning shell with the convective envelope.\\ 
We conclude that in our models thermohaline mixing during the 
HB changes the surface abundances only in stars with M~$< 1.5\mso$. 

\subsection{Asymptotic giant branch}\label{agb}

The subsequent evolutionary phase is characterized by two burning shells and a degenerate core.
The star burns H in a shell and the ashes of this process feed an underlying He-burning shell. This is referred to as the asymptotic giant branch (AGB) phase. 

During the low-luminosity part of the AGB thermohaline mixing works under the same conditions present in the last part of the HB phase (see Fig.~\ref{1.0}, label AGB).
In 1.0$\mso$ and 1.5$\mso$ models,  thermohaline mixing connects the shell source to the envelope. As a consequence surface abundances  change,  as shown for our 1.0$\mso$ 
model in Fig.~\ref{surfalles} (label AGB). Similarly to the RGB and HB phases, no thermohaline mixing is present in models with an initial mass higher than $1.5\mso$.

During the most luminous part of the AGB the He shell periodically experiences thermal pulses (TPs); in stars more massive
than $\sim2\mso$ these thermal pulses are associated with a deep penetration of the convective envelope, the so-called third dredge-up (3DUP).
In our $1\mso$ model we find thermohaline mixing to be present also in the TP-AGB.
The instability propagates  through the thin radiative buffer region (``thin'' in mass coordinates), and reaches the convective envelope. This situation is illustrated in Fig.~\ref{thpulse}.
But there thermohaline mixing only leads to negligible changes in the surface abundances. This because of to the very short timescale of this evolutionary stage and because most of the $\he3$ has already been burned in previous evolutionary phases. Overall in our models we found no impact of thermohaline mixing on the surface abundances of $\he3$ and on the $\c1213$ ratio during the TP-AGB phase. On the other hand thermohaline mixing can affect the surface abundance of lithium, as we discuss in Sect.~\ref{lithium_ab}.

We want to stress here that the presence and impact on surface abundances of thermohaline mixing during the TP-AGB, critically  depends on the local $\he3$ abundance and on the value of the efficiency factor  $\ath$.
This is because the local $\he3$ abundance is related to the previous history of mixing, which in turn also depends on the efficiency $\ath$ of the diffusion process.\\
We do not know the correct value of $\ath$ in stellar interiors. Indeed $\ath$ could also depend on stellar parameters such as rotation, metallicity or magnetic fields (see Sect.~\ref{rotation}), and it could well 
be that it changes in the same star through different evolutionary phases. Therefore our predictions for the changes of surface abundances due to thermohaline mixing, especially during the TP-AGB phase, are strongly affected by these uncertainties. 
Further study is needed to clarify the picture.

\section{Other mixing processes}\label{rotation}
\subsection{Other mixing processes in our models}
\begin{figure}
\resizebox{\hsize}{!}{\includegraphics{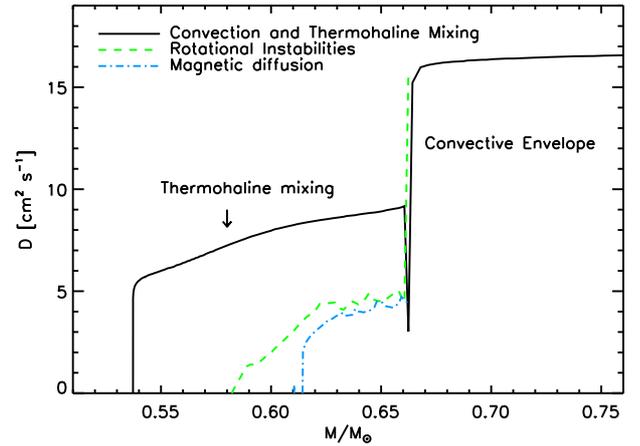}}
 \caption{Diffusion coefficients in the region between the H-burning shell and the convective envelope 
for the $1.0\mso$ model during the HB (t$=1.267\times10^{10}$). The initial equatorial velocity of
the model is $10 \kms$. 
The black, continuous line shows convective and thermohaline mixing diffusion coefficients, the green, 
dashed line is the sum of the diffusion coefficients due to rotational 
instabilities, while the blue, dot-dashed line shows the magnitude of the magnetic diffusion coefficient.}
\label{diffusion} 
\end{figure}

In our $1.0\mso$ and $1.5\mso$ models we found that in the relevant layers thermohaline mixing
has generally higher diffusion coefficients than rotational instabilities and magnetic diffusion. 
Figure \ref{diffusion} clearly shows that rotational and magnetically induced chemical diffusion is negligible compared to
the thermohaline mixing in our $1.0\mso$ model. The only rotational instability acting on a shorter
timescale is the dynamical shear instability, visible in  Fig.~\ref{diffusion} as a spike at  the lower 
boundary of the convective envelope.
This instability works on the dynamical timescale in regions of a star where a high degree of differential rotation is present; it 
sets in if the energy that can be gained from the shear flow becomes comparable to the work which has to be done against the
potential for an adiabatic turn-over of a mass element (``eddy'') \citep{heg98}. However, if present, this instability acts only in a 
very small region (in mass coordinates) at the bottom of the convective envelope. As a result thermohaline mixing is still setting the
timescale for the diffusion of chemical species from the convective envelope to the H-burning shell.

\begin{figure}
\resizebox{\hsize}{!}{\includegraphics{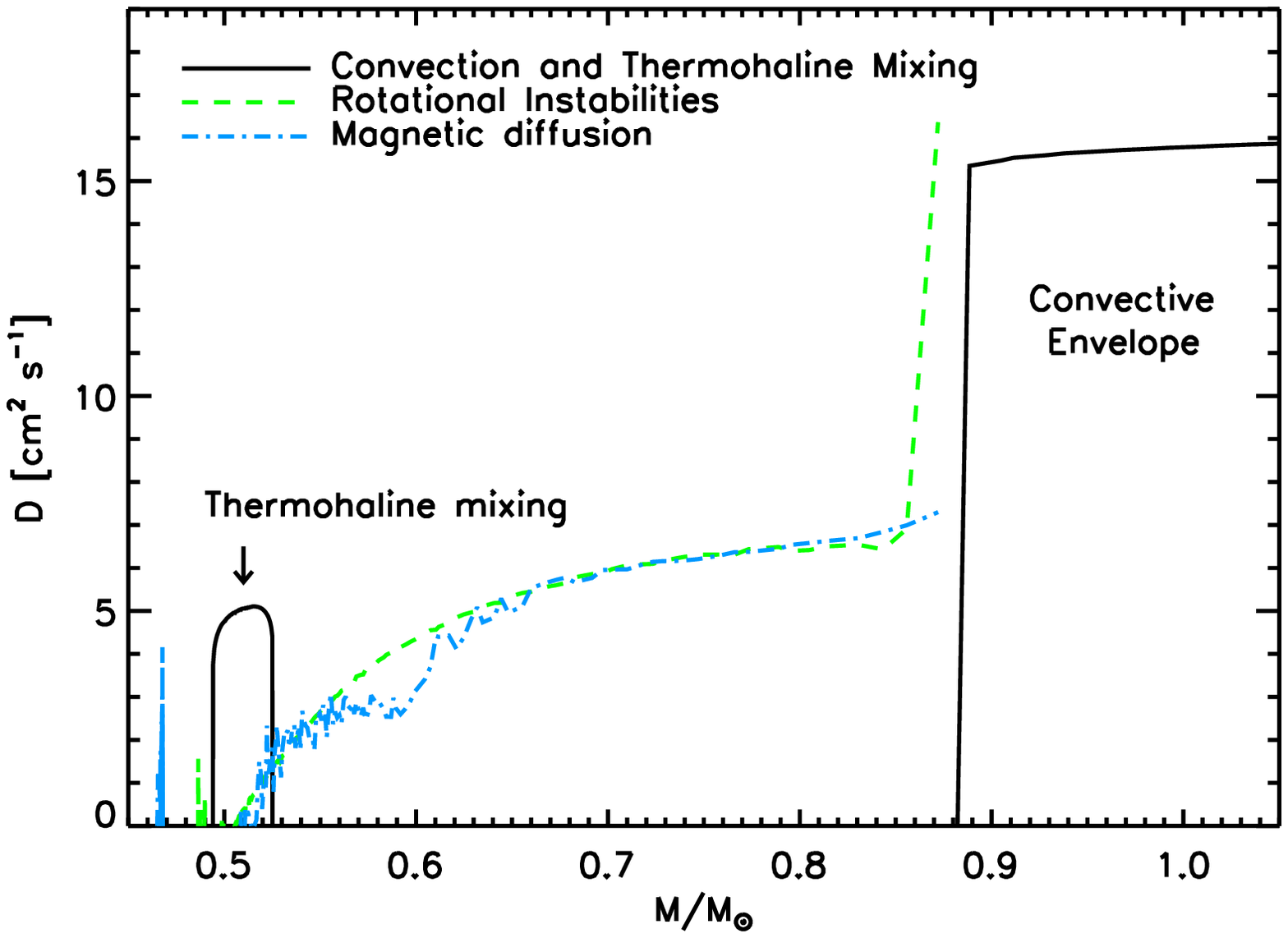}}
 \caption{Diffusion coefficients in the region between the H burning shell and the convective envelope 
for the $2.0\mso$ model during core He-burning (t$=1.124\times10^{9}$). The initial equatorial velocity of
the model is $140 \kms$. 
The black, continuous line shows convective and thermohaline mixing diffusion coefficients, the green, 
dashed line is the sum of the diffusion coefficients due to rotational 
instabilities while the blue, dot-dashed line shows the magnitude of magnetic diffusion coefficient.}
\label{diff2} 
\end{figure}

\begin{figure}
\resizebox{\hsize}{!}{\includegraphics{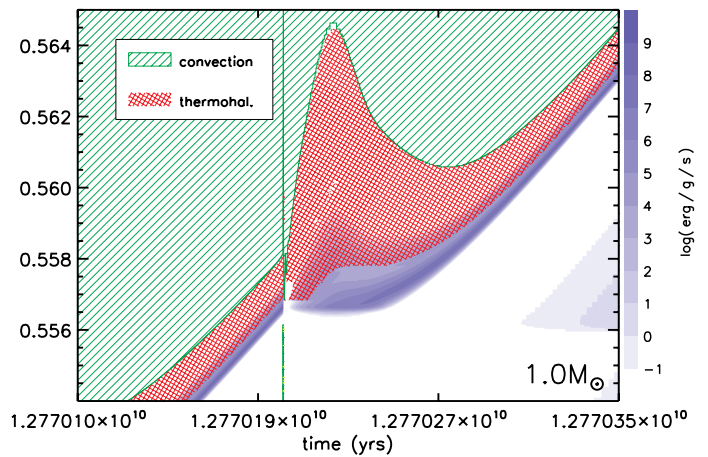}}
 \caption{Evolution of the region between the H-burning shell source and the convective 
envelope during a thermal pulse in a $1.0\mso$ star. Green hatched regions 
indicate convection, and regions of thermohaline 
mixing are red-cross hatched, as  displayed in the legend. Blue shading shows regions of 
nuclear energy generation. This model is evolved from the zero-age main sequence to the TP-AGB with $\ath = $2.}
\label{thpulse} 
\end{figure}

In models of  $2.0\mso$ and $3.0\mso$ thermohaline mixing is less efficient due to the lower abundance of $^3$He. At the same time rotational instabilities 
and magnetic diffusion have bigger diffusion coefficients, mainly because these models have initial equatorial velocities of 140 and 250 $\kms$ respectively. 
 Figure~\ref{diff2} shows how during core He-burning rotational mixing and magnetic diffusion become more important than thermohaline
 mixing in the $2 \mso$ model. The radiative buffer layer is dominated by the Eddington-Sweet circulation, dynamical shear, and magnetic diffusion. Yet the rotational mixing 
diffusion coefficient is still too small to allow the surface abundances to change appreciably in this phase, in agreement with results from \citet{Pct+06}. The same conclusion
is valid for the magnetic diffusion, which has the same order of magnitude as the rotational diffusion in the radiative buffer layer. 
Our models are calculated with the \citet{Kippenhahn:1980p5400} prescription for thermohaline mixing, which implies a smaller diffusion coefficient with respect to that proposed by \citet{Ulr72}.
As a consequence the result that thermohaline mixing has in general a higher impact than rotational mixing and magnetic diffusion in the relevant layers is valid regardless 
of which of the two prescriptions was chosen.

\subsection{Critical model ingredients}
The results described above are obtained with a particular model for rotational mixing and angular momentum transport, for which several assumptions need to be made.
Here we discuss the two most important assumptions in the present context.
The first assumption is that angular momentum transport in convection zones can be described with a diffusion approximation and a diffusion coefficient derived from the mixing length theory 
(Sect.~\ref{method}). The result is near rigid rotation in convection zones. Recent 3D hydrodynamic models of rotating red giant convective zones 
\citep{2007AN....328.1054S,2009ApJ...702.1078B} indicate that the picture may be more complex. \citet{2009ApJ...702.1078B} find radial profiles intermediate between constant angular momentum and constant angular velocity, with a large dependence on the rotation rate of the star.
While no general conclusion can easily be drawn from these studies, we may wonder how a reduced angular momentum transport efficiency in the convective envelope might affect our results. 
While detailed models would be required to exhaustively answer this question, we can expect that a more rapidly rotating base of the convective envelope would lead to less shear, and would thus render 
shear mixing in the layers below the envelope less relevant.

A second crucial assumption is the adoption of magnetic angular momentum transport according to \citet{Spruit:2002p36}. Even though the Spruit-Taylor dynamo has been criticized  
\citep{2007ApJ...655.1157D,2007A&A...474..145Z}, an effecient angular momentum transport mechanism like the Spruit-Taylor dynamo is clearly needed to understand the observed 
slow rotation of stellar remnants  \citep{hws05,2008A&A...481L..87S}. While angular momentum transport through gravity waves has been advocated as an interesting alternative  
\citep{2003A&A...405.1025T,2008A&A...482..597T}, it remains to be demonstrated that this  mechanism can work to break the rotation of the helium core in the post-main sequence 
stages of stellar evolution. Therefore, while the reader should be aware of the related uncertainties, using the Spruit-Taylor dynamo at this time appears reasonable.

\subsection{Interaction of instabilities}\label{interaction}
The discussion of the interactions of thermohaline 
motions with the rotational instabilities and magnetic fields is  complex.
In this respect \citet{Canuto:1999p5353} argues that shear due to differential rotation decreases the efficiency of thermohaline mixing. 
Not only \citet{Denissenkov:2008p5212} claim that rotation-induced horizontal turbulent diffusion may suppress thermohaline
mixing. This is because horizontal diffusion (molecular plus turbulent) may change the mean molecular weight of the fluid element during its motion. 
They argue that this horizontal diffusion is able to halt thermohaline mixing.
We think this argument is  correct in an ideal situation, in which a single blob of material is crossing an infinite, parallel slab.
Yet in a star the horizontal turbulence is acting on a shell, which can be locally approximated to a parallel slab with periodic boundary conditions in the horizontal direction.
This horizontal layer (shell) is rapidly homogenized by the horizontal turbulence. Fingers trying to cross this horizontal layer are quickly 
disrupted and mixed. This results in a rapid increase of the mean molecular weight $\mu$ in the shell, so that the region will become unstable to thermohaline mixing. A new generation of fingers is therefore expected. But the presence of horizontal turbulence is probably making fingers an unlikely geometrical configuration: blobs that travel a small distance before the turbulence is mixing them on a horizontal layer are more likely. This way thermohaline mixing is not stopped, but only slowed down. This scenario would favor the \citet{Kippenhahn:1980p5400} prescription, which actually predicts blobs traveling a distance comparable to their size.

Another interesting idea has been proposed by \citet{cz07b}. They claim that internal magnetic fields can play a stabilizing role, trying to counteract the destabilizing effect
of the inverse $\mu$ gradient. Their conclusion is that thermohaline mixing can be inhibited by a magnetic field stronger than $10^4-10^5$ Gauss.
But they warn that their analysis ignores both stellar rotation and the spatial variation of $B$, which results in neglecting any possible instability of the magnetic field 
itself \citep[e.g.,][]{Spr99}. 

The instability of magnetic fields below the  convective envelope of RGB and AGB stars has been discussed by \citet{Busso:2007p5186} and \citet{Nordhaus:2008p5185}. 
They argue that dynamo-produced buoyant magnetic fields could provide the source of extra mixing in these stars.

\section{Lithium-rich giants}\label{lithium_ab}
Lithium is a fragile element, which is destroyed at temperatures higher than about $3 \times 10^6$K. For this reason it is expected that lithium should decrease from its initial value during the evolution of stars. On the other hand, observations have shown that about $2\%$ of giants show strong Li lines \citep[e.g.,][]{Wallerstein:1982p5222,Brown:1989p5257}.
Some of these stars even show surface Li-abundances higher than the  interstellar values.

For intermediate mass stars a possible solution was proposed by \citet{Cameron:1971p517}, who showed how a net production of $^7$Li can be achieved  during hot-bottom burning (HBB).
During HBB the convective envelope penetrates into the H-shell burning, where $^7$Be is produced by the pp-chain. In this situation the unstable isotope $^7$Be  can be transported 
to cooler temperatures by the convective motions, decaying into $^7$Li in regions of the envelope where the temperature is low enough for lithium to survive.  This results in Li-enrichment at 
the surface.
   
At solar metallicity stars below $\sim 5\mso$ do not experience hot-bottom burning \citep{Forestini:1997p5288}, whereas at $Z=0$ hot-bottom burning is found down 
to $3\mso$ \citep{Siess:2002p5291}.  For stars avoiding hot-bottom burning, some other mechanism is needed in order to increase the Li surface-abundance. 
A possibility is that some kind of  extra mixing connects the H-burning shell and the convective envelope, which in the literature is often referred to as the cool bottom process (CBP).
%--------------------
The work of \citet{Charbonnel:2000p5307} supports this hypothesis. 
Indeed they found  Li-rich stars to be either red giants at the luminosity bump or early-AGB stars before the second dredge-up, in agreement with the idea that some internal mixing occurs when the H-burning shell enters a homogeneous region.  A lithium production during the RGB evolution is also supported by the recent observations of  \citet{2009A&A...508..289G}, who find lithium 
enriched red giants at the luminosity bump or at higher luminosities.

\citet{ulp07} reported the detection of low-mass, Li-rich AGB stars in the galactic bulge. Interestingly two of the four stars which show surface-Li enhancement present no evidence for third dredge-up, and thermohaline mixing is advocated as a possible source for the extra mixing.

\begin{figure}
\resizebox{\hsize}{!}{\includegraphics{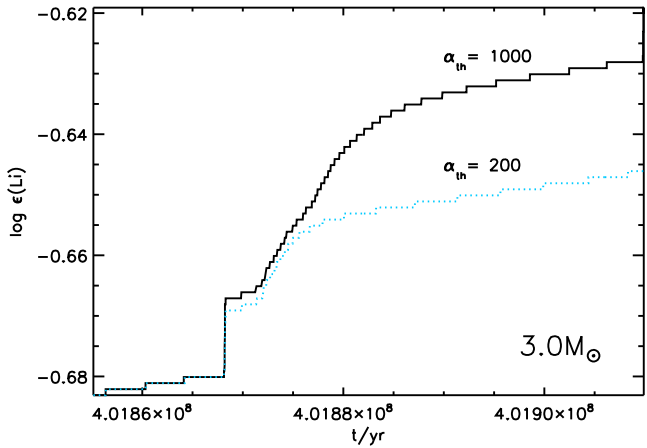}}
 \caption{Evolution of Li surface-abundance during one thermal pulse in a 3$\mso$ model. The black, 
continuous line shows a model evolved with $\ath$ = 1000; the blue, dotted line refers to the same model 
evolved with a thermohaline mixing efficiency $\ath$ = 200.
 In both cases the model  experiences third dredge-up. The evolution of the star prior to the TP-AGB has been 
calculated with $\ath$ = 2.}
\label{lithium} 
\end{figure}

In our calculations we found that the Li surface-abundance is affected by thermohaline mixing during the evolution of low-mass stars.
While Li is burned during the RGB and HB, thermohaline mixing has the potential to enhance the Li surface-abundance  during the TP-AGB phase. 
To show this, we computed stellar evolution calculations of the TP-AGB phase in 1 and 3 $\mso$ with different values of $\ath$. An example of the evolution of the Li surface-abundance in the 3$\mso$ model during one thermal pulse is shown in Fig.~\ref{lithium}.  Our models qualitatively confirm that this instability can enhance the surface Li abundances in low-mass AGB stars, even if we can not quantitatively reproduce  the high level of enrichment observed by \citet{ulp07}.  To reach the values of \citet{ulp07}  a value of $\ath$ much higher than those proposed by \citet{Kippenhahn:1980p5400} and \citet{Ulr72} is needed.  
As discussed in Sect.~\ref{agb}, a quantitative study requires a better knowledge of the efficiency parameter for thermohaline mixing $\ath$.

The observations of \citet{ulp07}  show that only 4 out of 27 galactic bulge stars are Li-enriched. 
If thermohaline mixing is the physical process providing the high Li-enrichment observed, we still have to understand why only 15\% of the sample show this strong 
enhancement. One possibility is  that these stars did not experience thermohaline mixing in previous evolutionary phases. This would leave  the $\he3$ reservoir intact, leading 
to a much more efficient mixing during the TP-AGB phase.\\
 This scenario requires a way to prevent the extra mixing during the RGB and HB phases. \citet{cz07b} have proposed that strong magnetic fields stop thermohaline mixing in those red giants
stars that are the descendants of Ap stars.  They call these stars ``thermohaline deviant stars". \\
Because the fraction of Ap stars relative to A stars (5-10 \%), the number of red giants that seem to avoid the extra mixing ($\sim 4\%$) and the observed fraction of Li-enriched AGB stars (15\%) are similar, it 
may be possible that we are looking  at the same group of stars at different evolutionary stages. If this is the case, it remains to be understood why the process that inhibits the mixing during the RGB and HB phases is not at work during the AGB.

A further complication arises from the observations of \citet{Drake:2002p5345}, showing that the incidence of Li-rich giants is much higher among fast-rotating objects. They consider single-K giants and find that among rapid rotators ($\vsini \ge 8 \kms$) a very large proportion ($\sim50\%$) is Li-rich, in contrast with a very low proportion ($\sim 2\%$) of Li-rich stars among the much more common slowly rotating giants. Thermohaline mixing is not driven by rotational energy, and if any effect would be expected, it would be a lower efficiency of the mixing with increasing shear and horizontal turbulence \citep{Canuto:1999p5353,Denissenkov:2008p5212}. On the other hand, an increase in the mixing efficiency with the rotation rate is expected if the physical mechanism behind the extra mixing is magnetic buoyancy \citep{Busso:2007p5186,Nordhaus:2008p5185,Denissenkov:2009p5218}. In this case rotation is necessary to amplify the magnetic field  below the convective envelope.\\
Another possibility is that lithium has an external origin, resulting from accretion and ingestion of planets or a brown dwarf by an expanding red giant \citep[e.g.,][]{Siess:1999p5380,Siess:1999p5382}. Mass transfer or  wind accretion in a binary system is also a possible scenario. 

The far-IR excess, which is observed in all fast rotating, Li enriched giants, is another interesting piece of the puzzle \citep{Drake:2002p5345,Reddy:2005p5386}. While models in which some kind of accretion process occurs could explain the IR excess, the internal production of lithium cannot  reproduce these observations \citep[but see][]{Palacios:2001p5388}.
We refer to \citet{Drake:2002p5345} for an accurate review of the proposed mechanism for the formation of Li-rich giants.

\section{Conclusion} 

We qualitatively confirm the results of CL07: thermohaline mixing in low-mass giants is capable
of destroying large quantities of ${^3}$He, as well as decreasing the ratio $\c1213$.
Thermohaline mixing indeed starts when the H-burning
shell source moves into the chemically homogeneous layers established by the first dredge-up. 
At solar metallicity we find that this process is working only in stars with a mass below $1.5\mso$.
This result is sensitive to the choice of the $\ath$ parameter, which regulates the speed of thermohaline mixing.

Our models show further that thermohaline mixing remains important during core He-burning
and can also operate on the AGB --- including the termally-pulsing AGB stage.   
Depending on the efficiency of the mixing process, this can result in considerable lithium enrichment.

Our calculations show that in the relevant layers thermohaline mixing generally  has a higher diffusion coefficient than rotational instabilities
and magnetic diffusion. 
However, we cannot address the interaction of thermohaline motions with differential rotation and magnetic fields, for which hydrodynamic calculations are required.

In stellar evolution codes thermohaline mixing is implemented as a diffusive process. This process acts on a thermal timescale,
but the exact velocity of the motion depends on a parameter $\ath$. This parameter is related to the geometry of the fingers (or blobs)
displacing the stellar material and is still a matter of debate. The two widely used prescriptions have a parameter $\ath$ that differs by two orders of magnitude.
We used the \citet{Kippenhahn:1980p5400} prescription, even though we also investigated the effect of using different values of $\ath$
in a few calculations. \citet{cz07} used a much more efficient thermohaline mixing \citep{Ulr72}, justifying their choice on the basis of
laboratory experiments of thermohaline mixing performed in water, and on the observations of surface abundances of red giants. \\
But the physical conditions inside a star are very different from these laboratory experiments, which clearly cannot be used for a quantitative 
study of this hydrodynamic instability. Moreover it is not clear if thermohaline mixing is the only physical process responsible for the extra mixing, and therefore
it is not possible to calibrate its efficiency against the observations.  

We argue that is not possible at this 
stage to firmly identify thermohaline mixing as the cause of the observed surface abundances in low-mass giants (Gratton et al. 2000). 
In particular the long standing $\he3$ problem cannot be considered  as solved.

In agreement with CZ07 we claim that to clarify the picture it would be desirable to have realistic hydrodynamic simulations of thermohaline mixing.

\begin{acknowledgements} 
MC  thanks Steve N. Shore, Maria Lugaro, Onno Pols, Evert Glebbeek, Selma de Mink, Jonathan Braithwaite, Miro Moc\'{a}k and John Lattanzio for helpful discussions.
MC acknowledges support from the International Astronomical Union and from the Leids Kerkhoven-Bosscha Fonds.
\end{acknowledgements}

\begin{appendix}
\section{how to stop thermohaline mixing}
For models of $\textrm{M} > 1.5 \mso$ we found two situations in which thermohaline mixing fails to connect the H-burning shell with the convective envelope:
\begin{enumerate}
\item The envelope is receding in mass coordinates and the thermohaline mixing is not fast enough to catch-up. This situation is shown in  Fig.~\ref{disconnected}.  
\item A chain of reactions rising the mean molecular weight can create a barrier that stops the mixing process. 
\end{enumerate}  
The second scenario occurs depending on the efficiency of the reactions rising the molecular weight in the outer wing of the H-burning shell. 
The reactions responsible for creating this compositional barrier are  $^3$He\,($^4$He,$\gamma$)\,$^7$Be\,(e$^-$,$\nu$)\,$^7$Li\,($^1$H,$^4$He)\,$^4$He. \\
In the first one $^3$He  and $^4$He produce $^7$Be 
that rapidly decays into $^7$Li. The lithium easily reacts with a proton producing two $\alpha$-particles; this way  the initial molecular weight of $8/9$ rises to the value $4/3$. 
This occurs in the outer wing of the H-burning shell, at a lower temperature with respect
to the region where the mean molecular weight inversion discussed by EDL06  and CZ07  is present.  
As a consequence thermohaline mixing is halted (see Figs.~\ref{2.0}  and \ref{peaks}).

Therefore two sets of reactions play a role in the evolution of thermohaline mixing in the region between the H-burning
shell and the convective envelope:

\begin{equation}
\left. \begin{array}{cc}
^3\mathrm{He}+^3\mathrm{He}   \longrightarrow 2\,^1\mathrm{H} + ^4\mathrm{He}  
\end{array} \right.
\;\;\mu: 1 \rightarrow 6/7 \label{he3he3} 
\end{equation}

\begin{equation}
\left. \begin{array}{cc}
^3\mathrm{He}+^4\mathrm{He}  \longrightarrow ^7\mathrm{Be}+\gamma \\
^7\mathrm{Be}+\textrm{e}^- \longrightarrow ^7\mathrm{Li}+\nu  \\
^7\mathrm{Li}  +^1\mathrm {H} \longrightarrow 2\,^4\mathrm{He}
\end{array} \right\}
\;\;\;\mu: 8/9 \rightarrow 4/3 \label{he3he4}.
\end{equation}
As a consequence, the efficiency in starting and stopping thermohaline mixing is regulated by the local abundance of $^3$He and $^4$He.
The first reaction in the chain \ref{he3he4} has a lower rate than reaction \ref{he3he3} 
at the temperatures  in the region of interest. However, $^3$He\,($^4$He,$\gamma$)\,$^7$Be depends linearly on the local abundance of $^3$He and $^4$He, 
while reaction \ref{he3he3} depends quadratically on the abundance of $^3$He. Increasing the initial mass of the stellar model, the ratio  $^3$He$/$$^4$He 
decreases in the radiative buffer layer \citep[e.g.,][]{Bs99}. Consequently if one increases the initial mass,  the set of reactions \ref{he3he4} at some point becomes more important than
\ref{he3he3}, meaning that a compositional barrier is created. This explains why we 
observe a threshold in mass at about $1.5\mso$, above which thermohaline mixing is not efficient anymore.\\ 

Note that both scenarios 1. and 2. depend on $\ath$, i.e. on the speed of the mixing process. Increasing the value of $\ath$ makes thermohaline mixing more difficult to be stopped; 
therefore the value of the mass threshold depends on the choice of $\ath$. 

\end{appendix}

\bibliographystyle{aa}

\end{document}